 \documentclass[12pt,preprint]{aastex}

\newcommand{\myemail}{shogo@z.phys.nagoya-u.ac.jp}

\shortauthors{S.Nishiyama et al.}

\begin{document}

\title{A Distinct Structure Inside the Galactic Bar}

\author{Shogo Nishiyama\altaffilmark{1,\bigstar}, 
Tetsuya Nagata\altaffilmark{2}, Daisuke Baba\altaffilmark{1},
Yasuaki Haba\altaffilmark{1}, Ryota Kadowaki\altaffilmark{1},
Daisuke Kato\altaffilmark{1}, Mikio Kurita\altaffilmark{1},
Chie Nagashima\altaffilmark{1}, 
Takahiro Nagayama\altaffilmark{2}, Yuka Murai\altaffilmark{2},
Yasushi Nakajima\altaffilmark{3}, Motohide Tamura\altaffilmark{3},
Hidehiko Nakaya\altaffilmark{4}, Koji Sugitani\altaffilmark{5},
Takahiro Naoi\altaffilmark{6}, Noriyuki Matsunaga\altaffilmark{7}, 
Toshihiko Tanab$\mathrm{\acute{e}}$\altaffilmark{7},
Nobuhiko Kusakabe\altaffilmark{8}, and Shuji Sato\altaffilmark{1}}

\altaffiltext{1}{Department of Astrophysics, Nagoya University, 
Nagoya, 464-8602, Japan}

\altaffiltext{$\bigstar$}{\myemail}

\altaffiltext{2}{Department of Astronomy, Kyoto University, 
Kyoto, 606-8502, Japan}

\altaffiltext{3}{National Astronomical Observatory of Japan, 
Mitaka, 181-8588, Japan}

\altaffiltext{4}{Subaru Telescope,  National Astronomical Observatory of
Japan, 650 North A'ohoku Place, Hilo, HI 96720, U.S.A.}

\altaffiltext{5}{Institute of Natural Science, Nagoya City University,
Nagoya, 464-8602, Japan}

\altaffiltext{6}{Department of Earth and Planetary Sciences The
University of Tokyo, Tokyo, 113-0033, Japan}
 
\altaffiltext{7}{Institute of Astronomy, School of Science, 
The University of Tokyo, Tokyo, 181-0015, Japan}

\altaffiltext{8}{Department of Astronomy and Earth Sciences,
Tokyo Gakugei University, Tokyo, 184-8501, Japan}

\begin{abstract}
We present the result of a near-infrared ($J H K_S$) survey
along the Galactic plane,
$-10\fdg5 \leq l \leq 10\fdg5$ and $b=+1\degr$,
with the IRSF 1.4m telescope and the SIRIUS camera.
$K_S$ vs. $H-K_S$ color-magnitude diagrams
reveal a well-defined population of red clump (RC) stars
whose apparent magnitude peak
changes continuously along the Galactic plane,
from $K_S=13.4$ at $l=-10\degr$ to $K_S=12.2$ at $l=10\degr$
after dereddening.
This variation can be explained by 
the bar-like structure found in previous studies,
but we find an additional
inner structure at $\mid l \mid\la 4^\circ$, 
where the longitude - apparent magnitude relation is distinct 
from the outer bar, and the apparent magnitude peak 
changes by only $\thickapprox$ 0.1 mag over the central $8^\circ$.  
The exact nature of this inner structure is as yet uncertain.

\end{abstract}


\section{Introduction}
\label{sec:intro}

The presence of the large-scale bar
in the central region of the Galaxy
was first suggested by \citet{deVau64}
and is now well established by a variety of methods:
gas dynamics \citep[e.g.,][]{Binney91}, 
bulge surface brightness \citep[e.g.,][]{Blitz91},
luminous star counts \citep[e.g.,][]{Nakada91},
bulge red clump stars \citep[e.g.,][]{Stanek94}, etc.
A consensus is slowly emerging for a large-scale Galactic bar with
an axis ratio of 3:1:1,
a major axis oriented at 15-40$^\circ$
with respect to the Sun-Galactic center line,
with the  nearer side at positive Galactic longitudes,
and a radius of more than 2.4 kpc \citep{Morris96}.

Smaller non-axisymmetric structures have been suggested
but no definite conclusion has been reached yet.
\citet{Alard01} made a star density map 
from 2MASS data
and suggested the existence of an inner bar
($\mid l \mid \lesssim 2\degr$ and 
$\mid b \mid \lesssim 2\degr$)
at an orientation different from the large-scale one.
\citet{Una98} counted the number of stars 
in the narrow $L$ band
at $b\thickapprox0\degr$, $l=\pm2.3\degr$ and $\pm4.3\degr$,
and found an asymmetry 
between positive and negative longitudes.
However, dereddened luminosity functions of point sources at 7$\mu$m
showed symmetric 
distributions
inside $\mid l \mid \lesssim 4\degr$
\citep{vanLoon03}.
Secondary bars are common
in external barred galaxies \citep[e.g.,][]{Laine02,Erwin04},
and they may play a crucial role 
in gas inflows to galactic centers.
Detailed observations of the inner region of the Galaxy
to clarify the spatial distribution of stars are thus essential.

To investigate the inner structure of the Galactic bulge,
we use red clump (RC) stars.
RC stars allow us 
accurate distance determinations
because of their narrow luminosity distribution
and large number in the bulge.
RC stars were employed by the OGLE group
\citep{Stanek94,Stanek96,Stanek97,Woz96,Pac98,Udal00},
and their observations
in the $V$ and $I$ bands provided 
clear evidence of the large-scale bar 
\citep{Stanek94,Stanek96}.
However, their observations were restricted to
a few windows of low extinction at high latitude 
($\mid b \mid \ga 2\degr$),
where data might not constrain 
the bar structure reliably \citep{Seven99}.

In this $Letter$, we present the results of near-infrared survey
at low Galactic latitude ($b=+1^\circ$).
Our results show clear evidence of 
an inner structure distinct from the large-scale bar.

\begin{figure}[b]
 \vspace{-0.5cm}
 \epsscale{.35}
 \rotatebox{-90}{
 \plotone{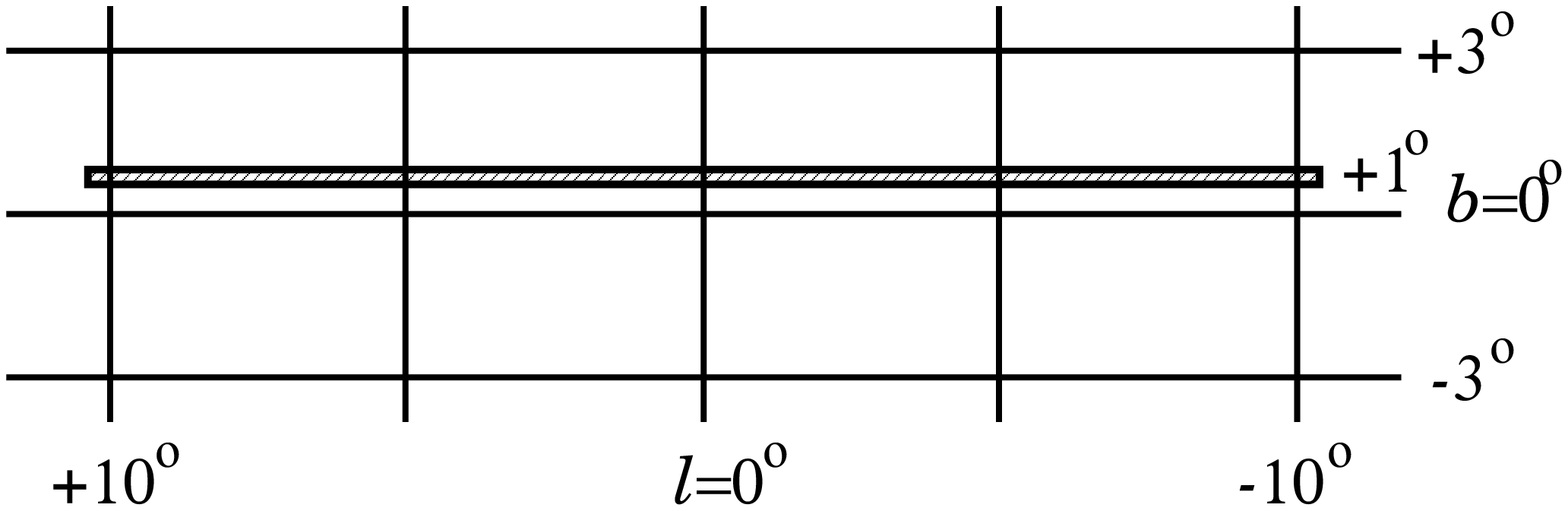}
 }
 \caption{Schematic map of the area observed,
from $l = -10 \fdg 5$ to $+10\fdg5$ at $b=+1^\circ$. 
 The width of the strip is about 8\arcmin.}
 \label{fig:obsarea}
\end{figure}

\section{Observations and Data Reduction}
\label{sec:obs}

\begin{figure}[t]
\epsscale{.80}
\plotone{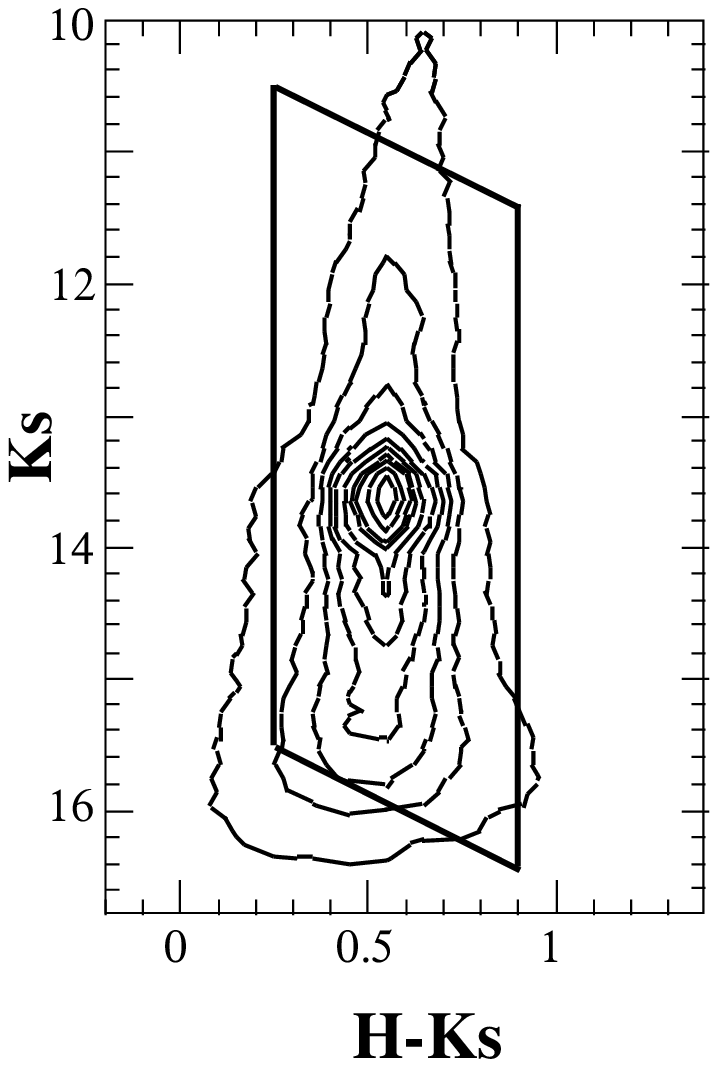}
\caption{$K_S$ vs. $H-K_S$ color-magnitude diagram 
obtained in our survey at $l=0\fdg0$. 
The parallelogram delineates
the region dominated by bulge red clump stars,
which in turn are used to make a $K_{H-K}$ histogram.
Contours are spaced linearly 
by 200 stars per $0.1^2$mag$^2$
between 100 to 1900 stars. }
 \label{fig:CMD}
\end{figure}

Our observations were obtained in 2004 May - June
using the near-infrared camera SIRIUS
\citep[Simultaneous InfraRed Imager for Unbiased Survey;][]{Nagas99,Nagay03}
on the IRSF (InfraRed Survey Facility) 
1.4m telescope of Nagoya University 
at South African Astronomical Observatory, South Africa.
SIRIUS can provide $J (1.25 \mu$m), $H (1.63 \mu$m),
and $K_S (2.14 \mu$m) images simultaneously,
with a field of view of 
$7\arcmin.7 \times 7\arcmin.7$ and
a pixel scale of $0\arcsec.45$.

At $b=+1^\circ$, we surveyed
a strip extending over
$\mid l\mid \leq 10\fdg5$, 
with a latitude width of about 8\arcmin (one field).
The total area surveyed is 
thus about 2.8 square degrees 
(see Fig. \ref{fig:obsarea}). 
Our observations were obtained only on photometric nights
and the typical seeing was 
1\farcs2 FWHM in the $H$ band.
A single image comprises 10 dithered 5 sec exposures, 
and the whole strip required 168 images.

\begin{figure}[ht]
 \epsscale{0.6}
 \plotone{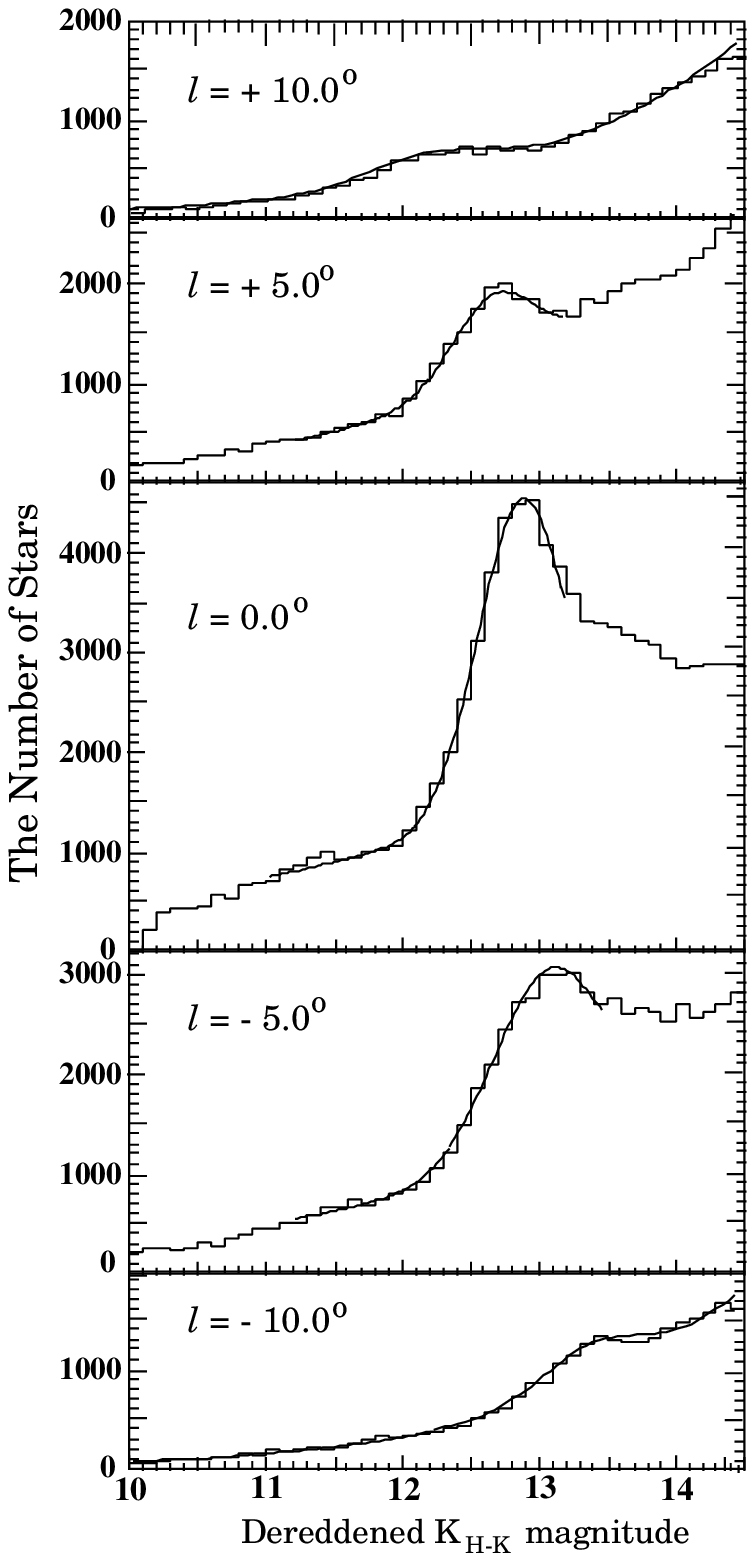}
 \caption{Histograms of the dereddened $K_{H-K}$
 magnitudes.
 Red clump stars make clear peaks 
 at
 $K_{H-K}\thickapprox 12.2, 12.7, 12.8, 13.0,$ and $13.4$ mag 
 at $l=+10\fdg0, +5\fdg0, 0\fdg0, -5\fdg0,$ and $-10\fdg0$, respectively. 
 Exponential and Gaussian functions are used to fit 
 the histograms (smooth line).}
 \label{fig:KHK}
\end{figure}

Data reduction was carried out with 
the IRAF (Imaging Reduction and Analysis Facility)\footnote{
IRAF is distributed by the National Optical Astronomy
Observatory, which is operated by the Association of Universities for
Research in Astronomy, Inc., under cooperative agreement with
the National Science.}
software package.
We applied standard procedures of 
near-infrared array image reduction 
in which dark frames were subtracted, flat-fields were divided, 
and sky frames were subtracted.
Photometry, including PSF fitting, 
was carried out with the DAOPHOT package 
\citep{Stetson87}.
We used the DAOFIND task to identify point sources
and the source lists were then input
for PSF-fitting photometry to the ALLSTAR task.
The PSF fit for each frame was constructed
with about 20 isolated stars.

Each frame was calibrated with the standard star
\#9172 in \cite{Persson98},
which was observed every 30 min.
We assumed that \#9172 is $H=12.12$ and $K_S=12.03$ 
in the IRSF/SIRIUS system.
The average of the zero-point uncertainties
was about 0.03 mag in both bands.
The 10$\sigma$ limiting magnitudes were
$H=16.6$ and $K_S=15.6$.

\section{Analysis and Results} 

\begin{figure}[t]
\epsscale{0.95}
 \plotone{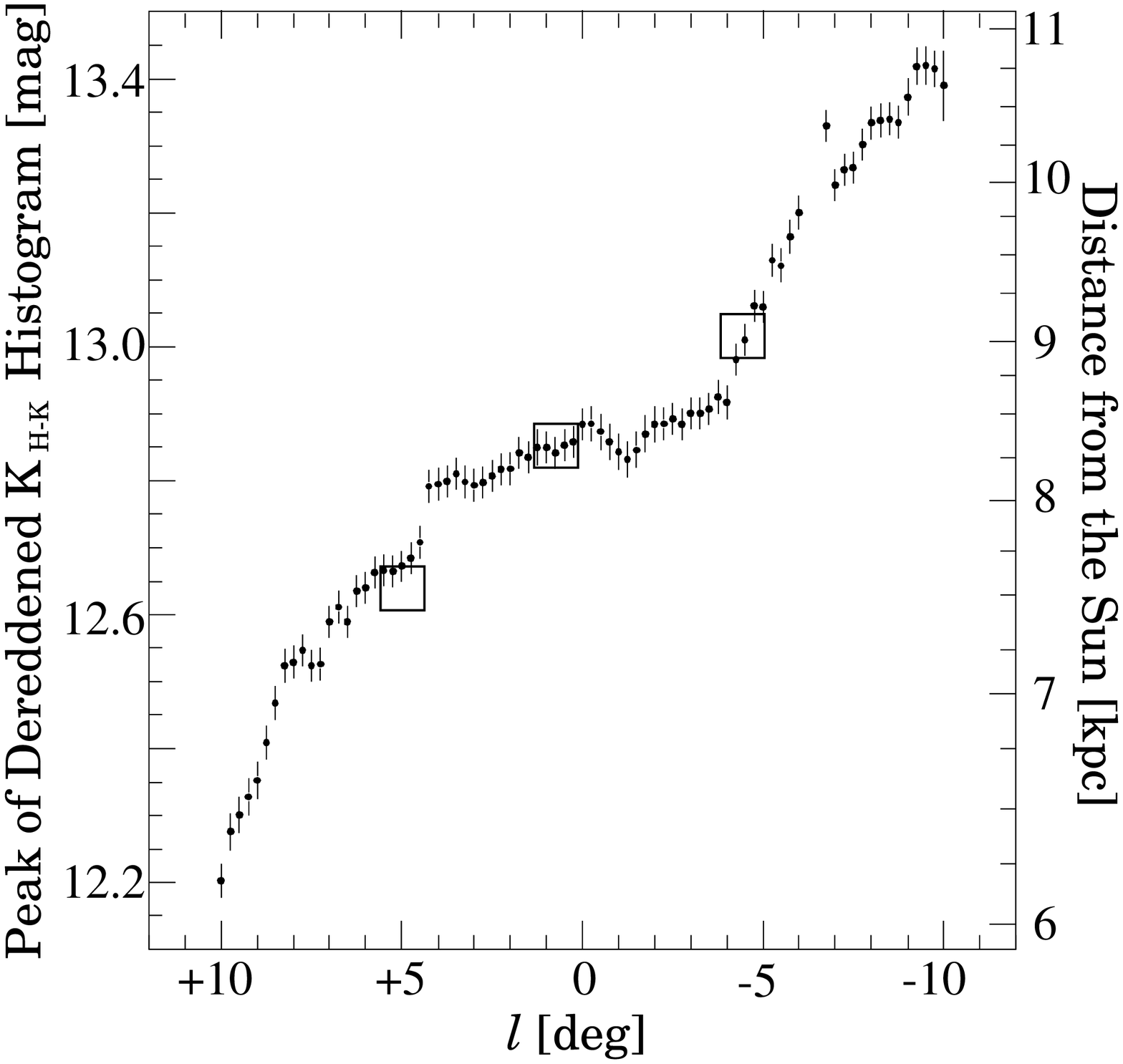}
 \caption{Dependence of the peak of 
 the $K_{H-K}$ distribution on the Galactic longitude $l$ 
 in our area surveyed (filled circle).
 The data of \citet{Stanek96} are also 
 plotted at $l \simeq -5,+1$ and $+5\degr$
 (open square).
}
 \label{fig:ang}
\end{figure}

RC stars are the equivalent of 
horizontal branch stars for a metal-rich population.
They constitute a compact and well-defined clump
in a color magnitude diagram (CMD)
and have a narrow luminosity distribution.
The mean magnitude of RC stars is thus a
good distance indicator,
and we apply the method developed by OGLE 
\citep{Stanek96,Stanek97} to our $H$ and $K_S $ data in this paper.

To analyze the distribution of RC stars, 
we define the extinction-free magnitude
\begin{displaymath}
 K_{H-K} \tbond K_S - \frac{A_{K_S}}{E(H-K_S)} \times \{(H-K_S)-(H-K_S)_{0}\}\,
\end{displaymath}
where we use the reddening law 
${A_{K_S}}/{E(H-K_S)} = 1.4$
\citep{Nishi05}\footnote{
Even if we adopt a different reddening law,
the resultant $K_{H-K}$ changes
only a few tenths mag,
and points in Fig. \ref{fig:ang}
move up or down 
as a whole with the {\it shape} unchanged.
}
and the intrinsic $H-K_S$ color of RC stars 
$(H-K_S)_{0} \simeq 0.1$ \citep{Alves00}.
$K_{H-K}$ is thus defined so that 
if ${A_{K_S}}/{E(H-K_S)}$ is independent of location,
then for any particular star 
$K_{H-K}$ is independent of
extinction.

First, we construct $K_S$ vs. $H-K_S$ CMDs
for individual regions of one degree in longitude.
For this, we merge the data of 8 
neighboring fields into one dataset,
e.g., we use the merged dataset 
of $ -0\fdg5 \leqq l \leqq +0\fdg5$ for $l=0\fdg0$,
$-0\fdg25 \leqq l \leqq +0\fdg75$ for $l=+0\fdg25$, etc.
Second, we extract the stars in the region 
 of the CMDs dominated by RC stars
(see, e.g., Fig. \ref{fig:CMD})
and the extracted stars are used in turn
to make $K_{H-K}$ histograms
(luminosity functions; see Fig. \ref{fig:KHK}).
The histograms have clear peaks
and are fitted with the sum of an exponential and a Gaussian functions
(see again, in Fig. \ref{fig:KHK}).
In addition to the RC peak,
a much smaller but clear peak also 
exists in the $\mid l \mid < 7\degr$ histograms
at a dereddened $K_{H-K}$ magnitude of $\thickapprox13.5$ 
(e.g., middle panel in Fig. \ref{fig:KHK}).
Fitting was made up to the magnitude
unaffected by this peak.
At $l=-6\fdg25$ and $-6\fdg50$,
the RC peaks appear distorted by this peak, 
so we excluded 
those datasets from our analysis.

Owing to highly nonuniform interstellar extinction 
over the region surveyed,
the peaks of RC stars in the CMDs shift
from one line of sight to another
over the range $13.0 \lesssim K_S \lesssim 14.5$. 
All the peaks are more than 1 mag brighter than 
the limiting magnitudes,
so they are well defined.
Also we confirmed a completeness 
of 89\% at $K_S=15$
by adding artificial stars into 
the most crowded frame ($l=+1\degr$).

The resulting dependence 
of the RC peak magnitude on the Galactic
longitude $l$ is shown in Fig. \ref{fig:ang}.
Error bars include uncertainties in the RC peak 
and the photometric calibration.
Assuming the absolute $K_S$ magnitude of RC stars 
to be $-1.61$ mag \citep{Alves00}, 
the absolute peak distances were also
calculated and are 
shown using the scale 
on the right of Fig. \ref{fig:ang}.
The distance to the Galactic center ($\mid l \mid < 1\degr$)
is $\thickapprox 8.3$ kpc,
slightly larger than the distance
of $8.0\pm 0.5$ kpc
obtained as a weighted average of 
a variety of observations \citep{Reid93}.

As shown by Fig. \ref{fig:ang},
the RC peak becomes 
progressively brighter at a greater galactic longitude,
from $K_{H-K}=13.4$ at $l=-10\degr$ to $K_{H-K}=12.2$ at $l=+10\degr$.
This is due to a nearer distance 
at $l>0\degr$ than $l<0\degr$
and clearly indicates that 
the bulge is not axisymmetric 
in support of previous work.
However, 
the slope of the points
in Fig. \ref{fig:ang} is not constant, 
becoming shallower at $\mid l \mid \la 4\degr$.
The peak magnitude changes by
only $\thickapprox 0.1$ mag 
over the central $8\degr$,
while it changes by more than 1 mag 
from $l=-10\degr$ to $l=+10\degr$.
This variation of the slope suggests 
the presence of an inner structure inside
the large-scale bar.

\section{Discussion}

RC stars are good distance indicators \citep{Alves00,Pietr03},
but their brightness weakly 
depends on their age and metallicity
\citep{Girar02,Salar2002,Groc2002}.
The bulk of stars in the Galactic bulge is 
very old \citep{Orto95,Zocc03},
and for such stars the variation in RC brightness
is estimated to be only 0.1 mag at $K$ 
for an age increase of 5 Gyr \citep{Salar2002}.
\citet{Frog99} and \citet{Ram00} did not find
any metallicity gradient 
along the Galactic major axis 
at $\mid l \mid \lesssim 4\degr$.
Although the gradient at larger $l$ 
along the Galactic plane 
has not been well studied,
no detectable gradient over the range 
$5\degr \lesssim l \lesssim 25\degr$ was found
by \citet[][]{Ibata95a,Ibata95b}.
Therefore our results are not likely to be affected
by age or metallicity variations of RC stars.

\citet{Stanek96} observed bulge RC stars 
in the $V$ and $I$ bands,
and estimated the modes of RC magnitudes.
They found that 
the modes are $\thickapprox 0.2$ mag brighter at $l = +5\degr$
and $\thickapprox 0.2$ mag fainter at $l = -5\degr$
than at $l \simeq +1\degr$. 
Their results are plotted in Fig. \ref{fig:ang}
so that the point 
at ($l,b) \simeq (+1\degr, -4\degr$)
agrees with our data.
The magnitude differences of the two datasets at $l=\pm 5\degr$
are in good agreement,
but the inner structure of the bulge escaped 
\citet{Stanek96}
because their observations were only 
in discrete low extinction regions.

The luminosity functions (LFs) of point sources at $7 \mu$m
made by \citet{vanLoon03}
from $l = -10\degr$ to $l = +10\degr$
at $\mid b \mid \leq 1\degr$ are similar
between positive and negative longitudes
for $\mid l \mid \leq 4\degr$.
Although their results imply that 
the inner region of the Galaxy is axisymmetric,
it would be very difficult to identify any
difference between the LFs
at the $\thickapprox 0.1$ mag level,
and the small asymmetry revealed by 
our observations might have been overlooked.
At a greater longitude, their LFs indicate that
the stars at $l\thickapprox +9\degr$ are generally closer to us
than at $l\thickapprox -9\degr$.
The large difference in apparent magnitude of RC stars
at $l\thickapprox \pm 9\degr$ in our work agrees with their results.

The narrow $L$ band observation of \citet{Una98}
did find an asymmetry in the star counts 
at $l=\pm2\fdg3$ and $l=\pm4\fdg3$.
Although their statistics 
were not very good,
they claim that the asymmetry is
best fitted with the interpolation of the models
developed by \citet{Blitz91} and \citet{Binney91,Binney97}
for the large-scale bar.
However we show here that 
the inner part makes a distinct structure,
so the interpolation might not be appropriate.
Non-axisymmetric structures 
even smaller ($\sim 1\degr$) than discussed here have also been suggested
\citep[e.g.,][]{Alard01,Sawada04},
and larger ($\ga 10-15\degr$) ones 
have also been discussed \citep[e.g.,][]{Lopez01},
but the scales of such structures are out of the range of our observation
and they are thus undetectable by our survey.

Recent studies of other galaxies suggest that
as many as one-third of all early-type barred galaxies
have secondary bars \citep[e.g.,][]{Laine02,Erwin04},
so the presence of an inner structure 
inside a large-scale bar is not uncommon.
Since the RC peak magnitude 
within $\mid l\mid \thickapprox 4\degr$ 
only slightly changes, 
the inner structure may well be a secondary bar
whose major axis makes a larger angle with respect to
the Sun - Galactic center line than the large-scale bar.
However, even if the real distribution of stars
in the inner part is axi-symmetric,
the observed RC peak magnitude 
could show a dependence
like that shown in Fig. \ref{fig:ang}
at $\mid l\mid \la 4\degr$.
In this region,
the stars of the inner structure and
those of the large-scale bar 
will overlap
along the line of sight,
yielding the observed shallow slope.
Although further modeling is thus clearly required to determine
the exact morphology 
of the inner structure,
our near-infrared observations have demonstrated
its existence towards the low Galactic latitude
of $b = 1\degr$.

\section{Summary}

In this paper,
we investigated the distribution of stars
in the Galactic center region
by using $H$ and $K_S$ data.
The longitude - apparent magnitude relation 
of red clump (RC) stars
in a thin strip along the Galactic plane
from $l=-10\fdg5$ to $l=+10\fdg5$ shows 
that the dereddened magnitude of the RC peak
changes continuously from $K_S=13.4$ to $K_S=12.2$.
This variation, which agrees well with optical data by OGLE,
supports the presence of the large-scale bar 
found in previous studies.
Most importantly, 
an inner structure distinct from the large-scale bar
is clearly found 
at $\mid l \mid\la 4^\circ$,
although its exact nature is still uncertain.

\acknowledgements

We would like to thank Yoshikazu Nakada and Martin Bureau
for their helpful comments.
We also thank the staff at SAAO for their support during our observations.
The IRSF/SIRIUS project was initiated and supported by Nagoya
University, National Astronomical Observatory of Japan
and the University of Tokyo in collaboration with 
South African Astronomical Observatory under 
Grants-in-Aid for Scientific Research
No.10147207, No.10147214, No.13573001, and No.15340061
of the Ministry of Education,
Culture, Sports, Science and Technology (MEXT) of Japan.
This work was also supported in part 
by the Grants-in-Aid for the 21st Century 
COE ``The Origin of the Universe and Matter: 
Physical Elucidation of the Cosmic History'' 
and ``Center for Diversity and Universality in Physics''
from the MEXT of Japan.

\end{document}